\documentclass[twocolumn,showpacs,amsmath,amssymb]{revtex4}
\usepackage{graphicx}
\usepackage{dcolumn}
\usepackage{bm}
\usepackage{amsmath}

\begin{document}

\title{Probing spin glasses with heuristic optimization algorithms}

\author{Olivier C. Martin}
\affiliation{LPTMS, Universit\'e Paris-Sud,\\
Orsay Cedex 91405, France}


\begin{abstract}

A sketch of the chapter appearing under the same
heading in the book ``New Optimization Algorithms in 
Physics''~\cite{HartmannRieger04} is given.
After a general introduction to spin glasses,
important aspects of heuristic algorithms for
tackling these systems are covered. 
Some open problems that one can hope to
resolve in the next few years are then considered.

\end{abstract}

\pacs{02.60.Pn  Numerical optimization; 
75.10.Nr Spin-glass and other random models}
\keywords{Spin glasses; combinatorial optimization; numerical optimization}

\maketitle

\section{Introduction}

Understanding the
physical properties of spin-glass materials is a long-standing
challenge~\cite{MezardParisi87b,FischerHertz91,Young98}.
Their exotic properties
are believed to be controlled 
by the low energy configurations, a configuration
being the specification of the 
orientation of each of the system's magnetic dipoles.
The physical origin of these dipoles is the ``spin''
of the electrons as in ordinary magnetism, but it is the
``irregular'' nature of the couplings between these
spins that gives these materials their striking properties.
To understand most of the phenomenology
of equilibrium spin glasses, it should be enough to find the
system's ground state
and low energy excitations: that is why much effort 
has focused recently on applying optimization algorithms
to spin glasses. Furthermore, spin glasses are
the archetypes of complex systems; what is learned here
is expected to have a strong impact on our understanding
of many other systems with competing effects.

In one of the first attempts to theoretically tackle
the competing interactions in
these systems, Edwards and Anderson~\cite{EdwardsAnderson75} (EA)
considered spins on a regular 
square or cubic lattice, and introduced spin-spin
couplings $J_{ij}$ as quenched independent random variables.
The corresponding Hamiltonian is
\begin{equation}
H_{\rm EA} = - \sum_{\langle i j \rangle } J_{ij} S_i S_j \ .
\label{eq:H_EA}
\end{equation}
They argued that the orientations of the spins $S_i=\pm1$ should
``order'' at low temperatures, but
at random, as would follow for instance from minimizing $H_{\rm EA}$.
Much of the theoretical research on spin glasses
since~\cite{BinderYoung86} has confirmed this idea, but it has been 
difficult to make firm claims accepted by all, to a large
extent because numerical computations are restricted to
small sizes and theoretical approximations are of
limited reliability. Work in this field has been quite fruitful
in the last 10 years; further progress may be expected
but it will require major advances in the algorithmics. We thus
consider now some of the corresponding issues.

\section{Using Heuristic Algorithms}

It is convenient to divide algorithms for finding ground
states into two classes: ``exact'' or complete, 
and ``heuristic'' or incomplete.
In the first class, the algorithms will terminate and for sure provide the
minimum energy solution; unfortunately they
tend to be slow.
Our focus here is on \emph{heuristic} algorithms: such algorithms
provide good but not necessarily optimal solutions. 
They should be used to find ground states only if one can 
measure their reliability, i.e., if can one convince oneself that
the failures to find
the true optima arise so rarely that they
make no difference for the study. Given this drawback,
heuristics have the advantage of being easy to
program and of being very fast; this often allows researchers
to quickly tackle relatively large systems with little pain.

\begin{figure}
  \begin{minipage}{8.0cm}
  \begin{minipage}{3.0cm}
  \hspace*{-0.45cm}\includegraphics[scale=0.165]{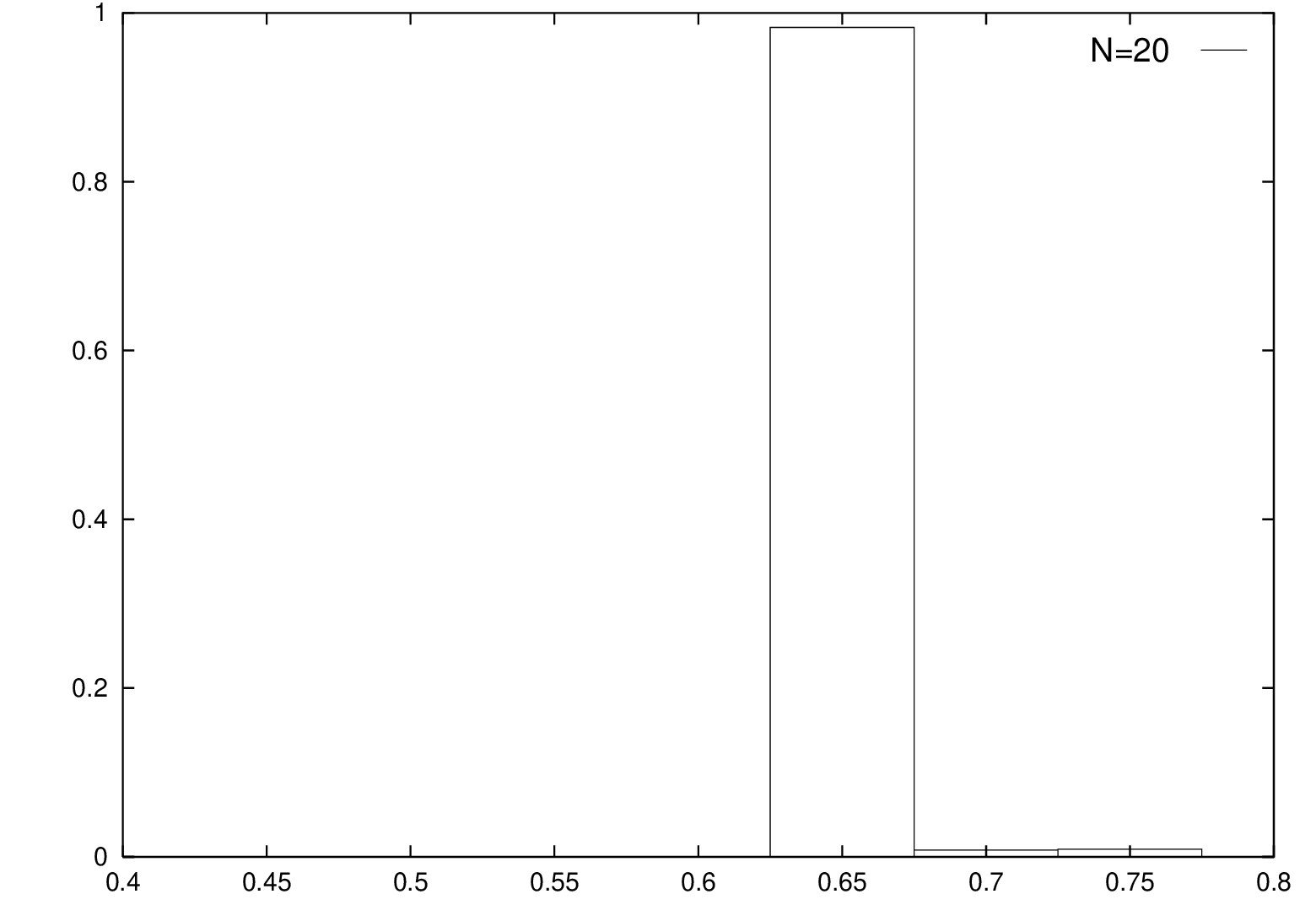}
  \end{minipage}
  \hfill
  \begin{minipage}{3.0cm}
  \hspace*{-0.95cm}\includegraphics[scale=0.165]{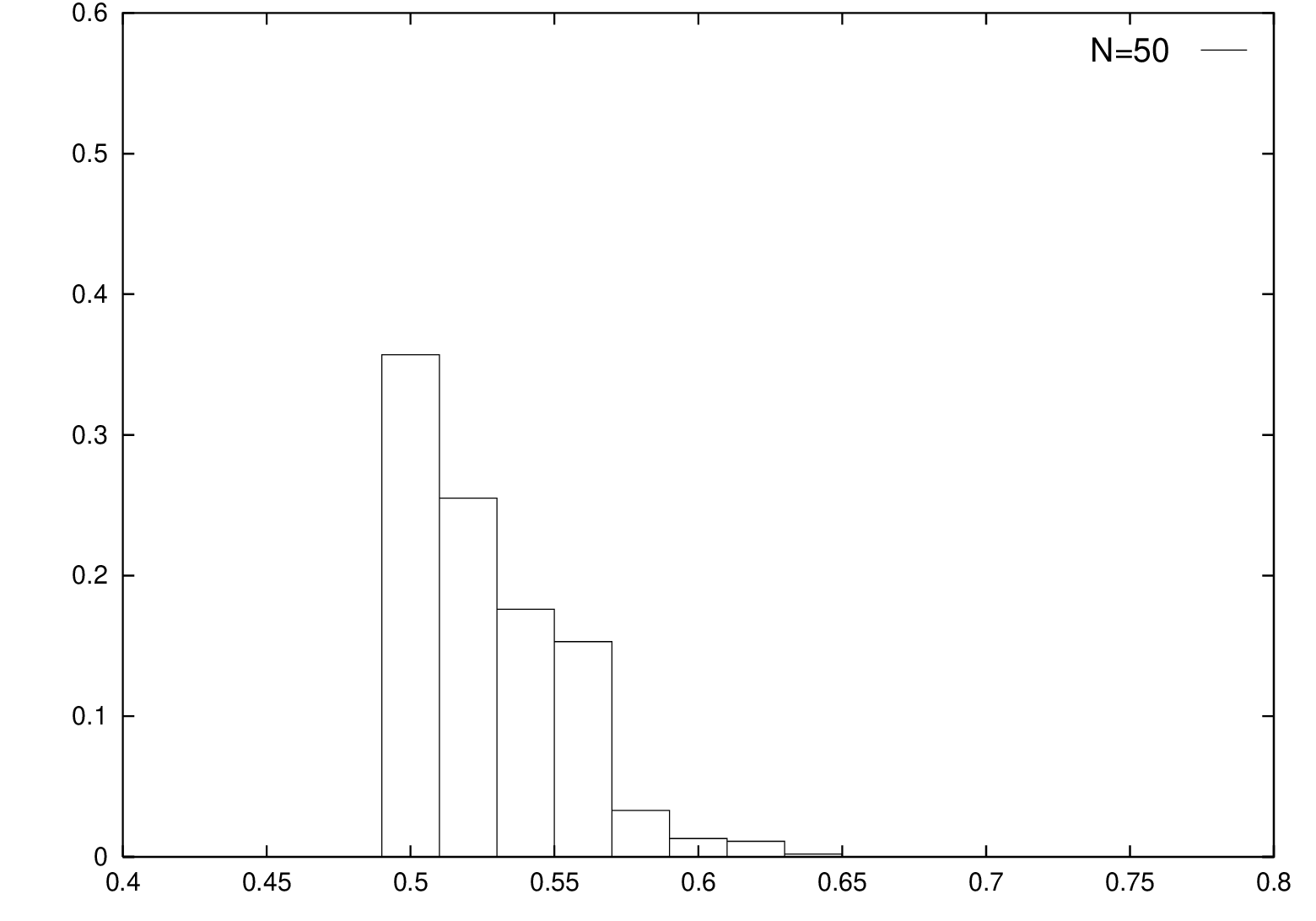}
  \end{minipage}
  \begin{minipage}{3.0cm}
  \hspace*{-0.45cm}\includegraphics[scale=0.165]{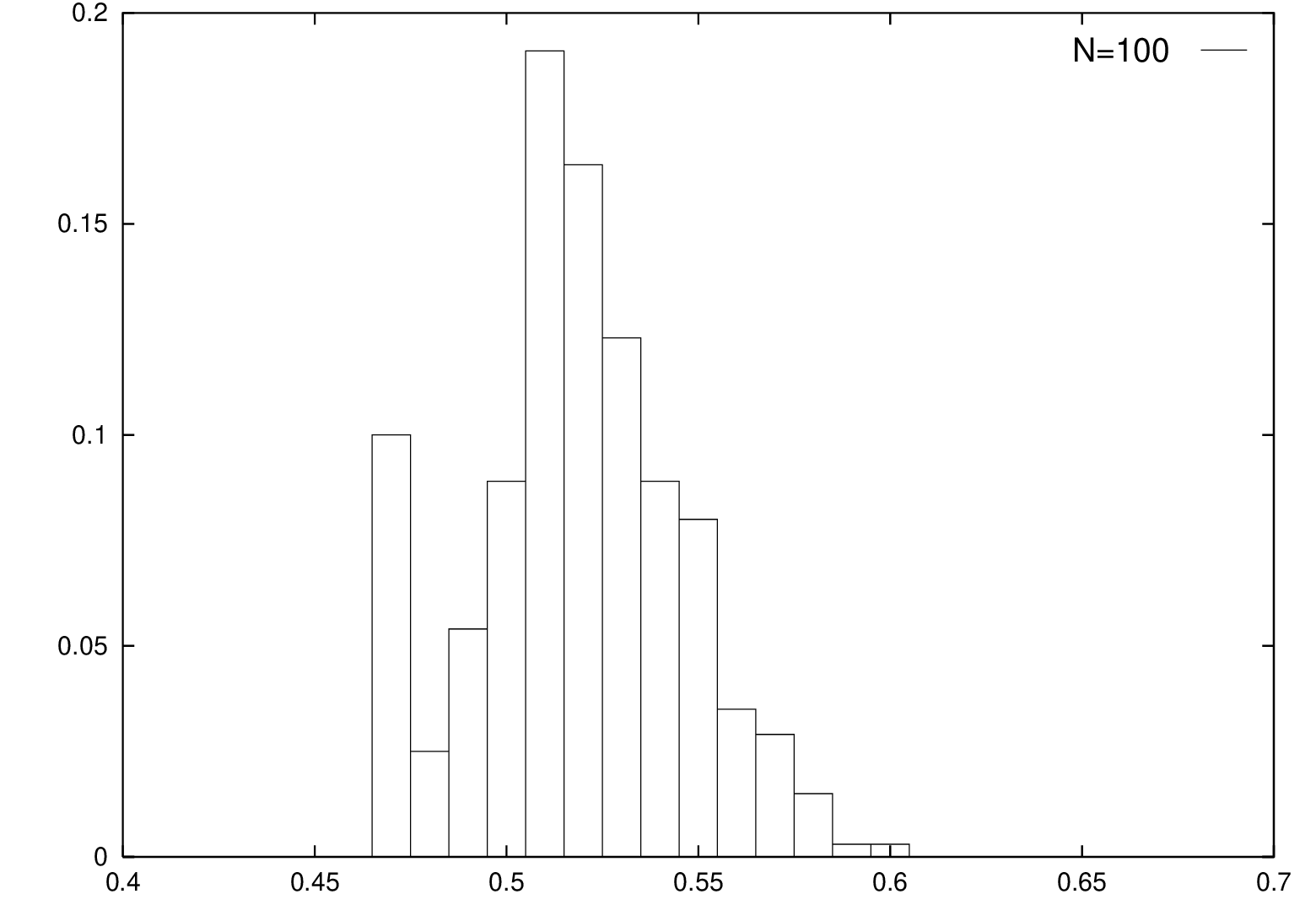}
  \end{minipage}
  \hfill
  \begin{minipage}{3.0cm}
  \hspace*{-0.95cm}\includegraphics[scale=0.165]{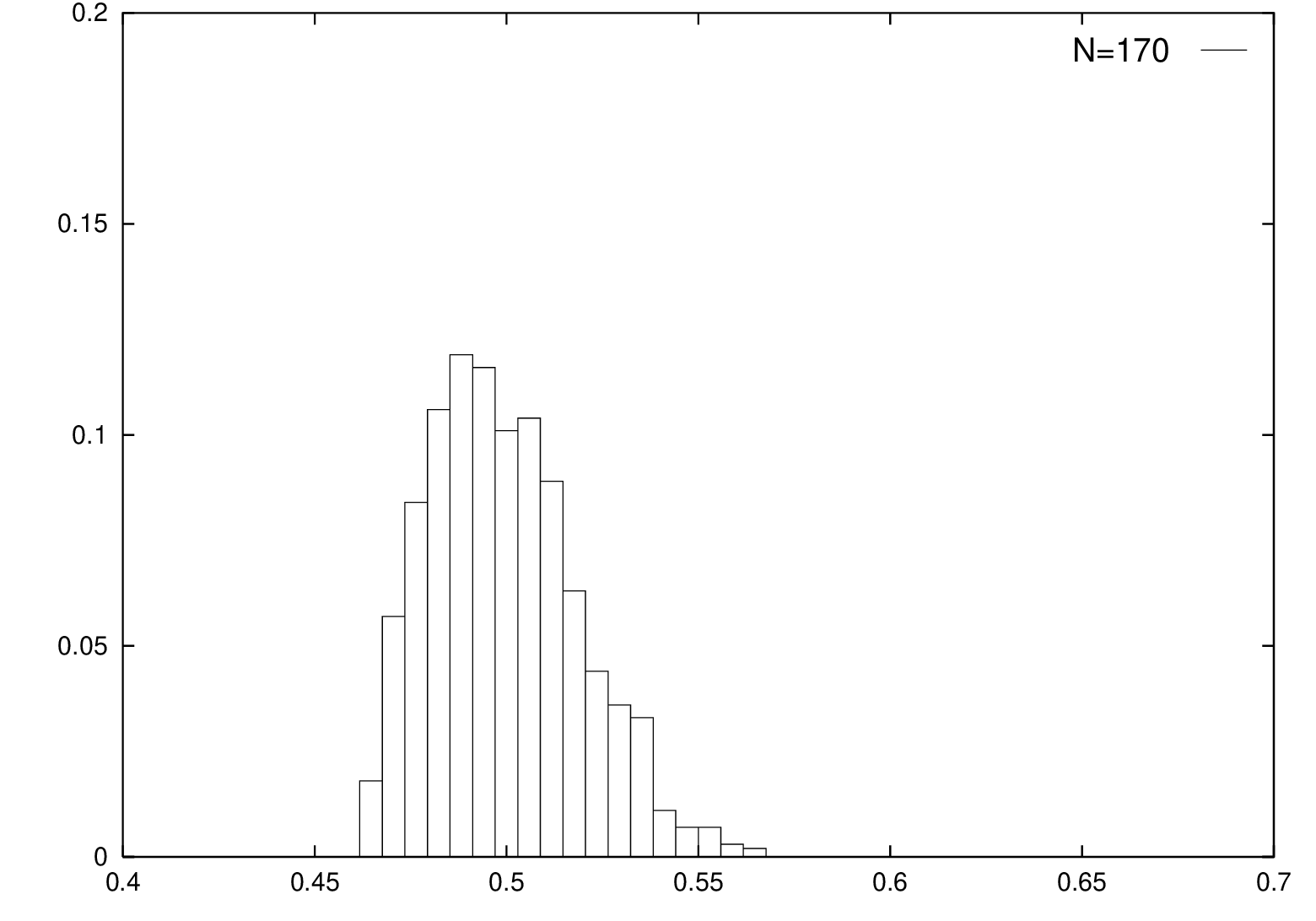}
  \end{minipage}
  \end{minipage}

  \caption{Distribution of energies per spin for the KL algorithm 
           applied to four samples with $N = 20, 50, 100$, and $170$ spins.}
  \label{fig:KLhisto}
  \end{figure}

If the Hamiltonian's ground state is unknown in general,
how do we measure the reliability of a heuristic algorithm?
We can perform a self-consistent test by running
{\it independent multiple starts} of
the algorithm. As an illustration, assume one has
performed $10$ independent runs of the algorithm. If all $10$
outputs are identical, there is good reason to expect that the
true ground state has been found. If on the contrary
the $10$ outputs do not all agree, the heuristic is
not so reliable. Generally, it seems unavoidable
for heuristics to become unreliable as the number
$N$ of spins becomes large; the challenge
is to postpone this bad behavior as much as possible.
To illustrate this point, we ran~\cite{SchreiberMartin99} 
a heuristic called
Kernighan-Lin~\cite{KernighanLin70}
on Ising spin-glass samples of increasing $N$. 
Fig.~\ref{fig:KLhisto}
shows the distributions of the energies per spin found
when using random starts: for each size ($N=20$, 50, 100 and 170),
the histogram is for a single sample, i.e., a single choice of the 
quenched disorder (the $J_{ij}$).

When $N=20$, 
the distribution is dominated by
the probability of finding the 
lowest energy $E_0$ (which presumably is that of the
ground state),
$P(E_0)\approx 0.98$; this problem size is thus extremely 
simple to ``solve''.
Increasing the instance size to $N=50$, the
distribution is still peaked at $E_0$ 
but much less so:
$P(E_0)\approx 0.35$. Going on to $N=100$, the distribution 
becomes bimodal.
Finally, for $N=170$, the peak at $E_0$ has
disappeared and the distribution is bell shaped.
All in all, as $N$ increases, the quality of the algorithm,
as measured for instance by $P(E_0)$,
deteriorates very clearly.
Any heuristic algorithm will follow this pattern of 
``simple'' to ``difficult'' as $N$ grows.
Such a behavior is generic;
the main difference from one heuristic algorithm to another is
the typical value of $N$ where there this
cross-over from simple to difficult arises.

In practice, a given heuristic will give estimates
of $E_0$ that are self-averaging, leading to
a fixed percentage error in the ground-state energy
density in the thermodynamic limit. Furthermore, 
the probability of finding the exact ground state 
tends to zero exponentially fast in the number of spins $N$.
These properties~\cite{SchreiberMartin99} 
allow one to compare different heuristics
and determine the usefulness of resorting to multiple
independent starts. Today's most used algorithms
find ground states with a high reliability
for 1000 spins but fail badly beyond 3000 spins.

\section{Some Physical Challenges}
Let's focus on the EA model in dimension 3 as the stakes there
are the highest. The definition of this model is very compact
(cf. Eq.~\ref{eq:H_EA});
it thus seems unbelievable
that even the {\em qualitative} features of this
system are still
subject to debate! Of interest to us here are the unsettled issues
directly related to optimization, namely
the nature of the ground state and its low-energy excitations.
In the last few years,
several teams~\cite{Hartmann97,HoudayerMartin99b,
PalassiniYoung99a,MarinariParisi00c,PalassiniLiers03},
each with their own codes, have investigated
these issues that we now highlight.

{\em Phase diagram ---}
In most studies, the $J_{ij}$ in Eq.~(\ref{eq:H_EA}) are taken
from a distribution symmetric about $0$. If instead
$J_{ij}<0$ for only a small fraction of the 
couplings, the system is ferromagnetic and
resembles the ordinary Ising model. As the fraction
of anti-ferromagnetic couplings is increased
beyond a critical value, the ferro-magnetism
disappears. Does the spontaneous magnetization
vanish just as the spin-glass
ordering appears? Can there be a ``mixed''
phase where the two types of orderings co-exist?
Such a phenomenon is expected within the
mean-field picture~\cite{MezardParisi87b}.
On the contrary, the droplet/scaling
picture~\cite{FisherHuse86,BrayMoore86}
says no such coexistence can arise. Only
recently has this question
been considered numerically
but the issue is still far from settled.

An analogous situation appears when
an external perturbation in the form of a magnetic
field $B$ is applied.
In the droplet/scaling picture, the spin-glass ordering is
destroyed as soon as $B \ne 0$, leading to a paramagnetic 
phase.
On the contrary, in the mean-field picture, spin-glass ordering
{\em co-exists} with the net magnetization induced by $B$
as long as $B$ is not too large. Numerical 
studies suggest that no co-existence occurs
in 3 dimensions, but further work is necessary.

{\em Energy landscapes ---}
Now consider the organization of the low-energy
configurations. Can one give
a statistical description whereby some of their
properties ``scale''? By that we mean
that for instance a characteristic energy has a 
power-law scaling with $N$ as $N \to \infty$. ($N$ is the number
of spins in the system.) The mean-field picture
predicts the following properties.

(i) Clustering: 
Given a low energy configuration, it is possible to further flip
1, 2, and more spins if they are carefully chosen 
without changing substantially the value of the
excitation energy. The
set of low energy configurations form families or ``clusters'';
two configurations belong to the same cluster if their
Hamming distance is very small compared to $N$.

(ii) Replica symmetry breaking:
Among the clusters formed by the low energy configurations, 
consider the ones in which a finite fraction $x$ of the
spins are flipped compared to the ground state. (By finite we mean
that $x$ is fixed, $0 < x < 1$, and then one focuses on
the large $N$ limit with that given $x$.)
In mean field, the corresponding
{\em lowest} excitation energy is $O(1)$.
Furthermore, the system has
{\em continuous} replica symmetry breaking, meaning that 
$x$ can take on values 
that span at least some sub-interval of $\left[ 0, 1 \right]$.

(iii) Ultrametricity: 
One defines the distance 
between two configurations as their
Hamming distance divided by $N$. One can also
define the distance $d(\alpha,\beta)$
between two {\em clusters} of configurations 
from the mean distance of their respective members.
In mean field, it turns out that 
the low energy clusters are organized hierarchically:
each cluster is divided into sub-clusters
which are themselves
sub-divided\ldots~ Furthermore, if we think of clusters
as being points in an abstract space,
that space is {\em ultrametric}, i.e., all triangles are 
isosceles.

It is a major challenge to understand whether these remarkable
properties also arise in the 3-dimensional EA model. 
Of course they may not; for instance
the droplet/scaling picture may hold instead. In that picture, 
scaling laws play a central role but so do
``position-space'' properties. At the heart of these is the notion
of \emph{droplet} excitations above the ground state; a droplet
is defined on a given scale $\ell$ and around 
a given spin $S_0$ as follows. One considers all connected
clusters of spins that are
enclosed in the cube of side $2 \ell$ centered on $S_0$ but which
are not contained in that of side $\ell$; among all these
potential excitations, the droplet is the one
of lowest energy. The authors of the
droplet/scaling picture postulate
the following properties. (i) Droplet energies grow as 
a positive power of their volume.
(ii) There is no replica symmetry breaking.
(iii) The organization of the low energy clusters
is not hierarchical; nevertheless, 
the energy landscape is self-similar, i.e., it is a fractal.

Just as for the phase diagram, studies of energy
landscapes have not led to any consensus in spite
of much work. The root of this problem
may very well be that new theoretical 
frameworks~\cite{KrzakalaMartin00,PalassiniYoung00a} are necessary.

\section{Outlook}
If spin glasses are so controversial, it is largely because
finite-size effects in these systems 
are not well understood. Without such
an understanding, it is very difficult to extrapolate
to the thermodynamic limit, i.e., $N \to \infty$.
This means that algorithmic improvements have to be important,
allowing us to go to significantly further in $N$.
Currently, we can perform studies involving about $2000$ spins.
It would probably be necessary to at least 
double that to have 
good enough control over the finite-size corrections
and to settle long-standing disputes ``beyond reasonable doubt''.

\bibliographystyle{prsty}
\bibliography{../../Bib/references,../../Bib/co}

\end{document}